\documentclass[american,aps,pra,reprint,superscriptaddress, longbibliography]{revtex4-1}
\usepackage[T1]{fontenc}
\usepackage[latin9]{inputenc}
\usepackage{color}
\usepackage{babel}
\usepackage{amsthm}
\usepackage{amsmath}
\usepackage{amssymb}
\usepackage{graphicx}
\usepackage[unicode=true,pdfusetitle, bookmarks=true,bookmarksnumbered=false,bookmarksopen=false,  breaklinks=false,pdfborder={0 0 0},backref=false,colorlinks=false] {hyperref}
\hypersetup{ colorlinks,linkcolor=myurlcolor,citecolor=myurlcolor,urlcolor=myurlcolor}
\usepackage{tcolorbox}

\makeatletter
\@ifundefined{textcolor}{}
{%
 \definecolor{BLACK}{gray}{0}
 \definecolor{WHITE}{gray}{1}
 \definecolor{RED}{rgb}{1,0,0}
 \definecolor{GREEN}{rgb}{0,1,0}
 \definecolor{BLUE}{rgb}{0,0,1}
 \definecolor{CYAN}{cmyk}{1,0,0,0}
 \definecolor{MAGENTA}{cmyk}{0,1,0,0}
 \definecolor{YELLOW}{cmyk}{0,0,1,0}
 }
\theoremstyle{plain}
\newtheorem{thm}{\protect\theoremname}
\theoremstyle{plain}

\ifx\proof\undefined
\newenvironment{proof}[1][\protect\proofname]{\par
\normalfont\topsep6\p@\@plus6\p@\relax
\trivlist
\itemindent\parindent
\item[\hskip\labelsep
\scshape
#1]\ignorespaces
}{%
\endtrivlist\@endpefalse
}
\providecommand{\proofname}{Proof}
\fi
\theoremstyle{plain}

\newtheorem{defi}[thm]{\protect\definitionname}

\providecommand{\lemmaname}{Lemma}
\providecommand{\definitionname}{Definition}
\providecommand{\propositionname}{Proposition}

\usepackage{babel}
\usepackage{txfonts}
\usepackage{colortbl}\definecolor{myurlcolor}{rgb}{0,0,0.7}
\newcommand{\tr}{{\operatorname{Tr\,}}}

\def\ket#1{| #1 \rangle}

\def\proj#1{| #1 \rangle\!\langle #1 |}
\newcommand{\ketbra}[2]{|#1\rangle\!\langle#2|}

\newcommand{\haH}

\definecolor{orange}{RGB}{255,127,0}

\begin{document}
\title{Quantum Bayes' Rule Affirms Consistency in Measurement Inferences in Quantum Mechanics
}

\author{Mohit Lal Bera}
\affiliation{ICFO -- Institut de Ciencies Fotoniques, The Barcelona Institute of Science and Technology, ES-08860 Castelldefels, Spain}

\author{Manabendra Nath Bera}
\email{mnbera@gmail.com}
\affiliation{Department of Physical Sciences, Indian Institute of Science Education and Research (IISER), Mohali, Punjab 140306, India}

\begin{abstract}
Classical Bayes' rule lays the foundation for the classical causal relation between cause (input) and effect (output). This causal relation is believed to be universally true for all physical processes. Here we show, on the contrary, that it is inadequate to establish correct correspondence between cause and effect in quantum mechanics. In fact, there are instances within the framework of quantum mechanics where the use of classical Bayes' rule leads to inconsistencies in quantum measurement inferences, such as Frauchiger-Renner's paradox \cite{Frauchiger18}. Similar inconsistency also appears in the context of Hardy's setup \cite{Hardy92, Hardy93} even after assuming quantum mechanics as a non-local theory. As a remedy, we introduce an input-output causal relation based on quantum Bayes' rule. It applies to general quantum processes even when a cause (or effect) is in coherent superposition with other causes (or effects), involves nonlocal correlations as allowed by quantum mechanics, and in the cases where causes belonging to one system induce effects in some other system as it happens in quantum measurement processes. This enables us to propose a resolution to the contradictions that appear in the context of Frauchiger-Renner's and Hardy's setups. Our results thereby affirm that quantum mechanics, equipped with quantum Bayes' rule, can indeed consistently explain the use of itself. 
\end{abstract}

\maketitle

\section{Introduction}
In any process, an observed effect (output) can be attributed to the cause combining the initial state (input) and the evolution it has undergone. In an arbitrary classical (stochastic) process, the classical causal relation (CCR) allows one to deterministically predict an effect for a given cause if the conditional probability (or transition probability encoding the evolution) is unity. Similarly, an observed effect can be used to infer a cause with certainty if the transition probability corresponding to the ``inverse'' process, also known as the retrodicted conditional probability, is unity. The transition probabilities for the ``retrodicted'' process are derived using classical Bayes' rule \cite{Jaynes03}. Say, an event (cause or input) $a$ occurs with probability $P(a)$ and undergoes a stochastic evolution to give rise to another event $b$ (effect or output) with probability $P(b)$. The conditional probability encoding the evolution is given by $P(b|a)$. It means, although $a$ occurs with probability $P(a)$, any occurrence of $a$ can predict observation of $b$ with the probability $P(b|a)$. To make an inference, i.e., finding out a cause of observing $b$, the process has to be inverted, and this is done using the Bayesian retrodiction rule given by
\begin{align*}
	P(a|b)=\frac{P(b|a) \  P(a)}{P(b)}.
\end{align*}   
It implies that whenever we observe the event $b$, we can definitely connect it to the cause $a$ with probability $P(a|b)$.

Traditionally, the CCR based on classical Bayes' rule is applied to all physical theories, including quantum mechanics. Although, quantum mechanics is fundamentally different from its classical analog in many aspects.  First, the former allows superpositions of orthogonal (perfectly distinguishable) states. It also allows superposition in evolutions \cite{MNBera19}. Thus, quantum mechanics allows a superposition of causes, that may represent a different cause, leading to a superposition of effects that can again be seen as a different effect. Second, there are non-deterministic processes, such as quantum measurements, where the system's original state collapses to some other state with a probability. Third, quantum mechanics allows nonlocal correlations that lead to ``spooky action at a distance'' \cite{EPR35}. As a consequence, for a correlated composite, a measurement made on one subsystem may induce collapse on the other. Can these peculiarities of quantum mechanics be captured by the CCR as the latter, a priory, does not take into account any such quantum features? It is now known that CCR leads to paradoxical results in quantum mechanics. One prominent example is Frauchiger-Renner's \cite{Frauchiger18} paradox. There the prediction based on a given past directly contradicts with the inferences made about the same past for observations made at the present.  Therefore, revisiting the applicability of CCR in the quantum domain is important and may find fundamental implications in the foundation of quantum mechanics.

Here we show, in contrast to common belief, that the classical Bayes' rule is inadequate to establish a consistent causal correspondence between quantum causes and effects. We introduce a quantum (input-output) causal relation (QCR) based on quantum Bayes' rule applicable to general quantum processes. With two examples, we demonstrate how classical and quantum causal relations lead to contradictory causal inferences for the same observations or effects and why CCR is inadequate. In addition, we provide conditions for deterministic causal correspondence between quantum cause and effects. The QCR accounts for the situation where a cause is in a superposition of other causes and, similarly, for the effects. Beyond that, it correctly describes the causal correspondence between causes belonging to one quantum system and effects belonging to the others for a global process. We propose a resolution to the contradictions that appear in the case of Frauchiger-Renner's paradox with the help of QCR. Furthermore, we revisit Hardy's setup leading to bipartite nonlocality without a Bell inequality \cite{Hardy92, Hardy93} and show that, within the framework of quantum mechanics (a non-local theory), there also appears contradiction in prediction and inferences. Yet again, it is resolved with the help of QCR. Thus our results advocate that, while deriving correct correspondence between cause and effect in quantum mechanics, one must resort to quantum Bayes' rule. 

The article is organized as follows. In section \ref{sec:CRs}, we briefly discuss quantum conditional states and quantum Bayes' rule and introduce deterministic quantum causal relations. In section \ref{sec:CCRvsQCR}, we consider quantum processes in which the predictions and inferences based on CCR and QCR drastically differ. Sections \ref{sec:FRvsQCR} and \ref{sec:HvsQCR} revisit Frauchiger-Renner's and Hardy's setups, respectively. We demonstrate that the contradictions that appear there are due to CCR, and these may be resolved by exploiting QCR. Finally, we conclude in section \ref{sec:Disc}.

\section{Quantum Conditional States, Bayes' rule, and Causal Relations \label{sec:CRs}}
For classical systems, the states are described by the probabilities, and (stochastic) processes are expressed in terms of conditional probabilities. However, for a quantum system, probabilities are not sufficient. One needs to express the states in terms of density matrices which, in addition to probabilities, carry the information about the quantum superposition they may have. For the same reason, conditional probabilities should be upgraded to conditional states capable of encoding the information about quantum evolution and possible superpositions in causes and effects. Below we define the conditional states, quantum Bayes' theorem, and provide conditions for deterministic causal relations for quantum processes.

Consider a quantum evolution by a completely positive trace preserving (CPTP) map $\Lambda: \mathcal{L}(\mathcal{H}_S) \mapsto \mathcal{L}(\mathcal{H}_R)$ where $\mathcal{H}_S$ and $\mathcal{H}_R$ are the corresponding Hilbert spaces of the systems $S$ and $R$ respectively. The quantum causal conditional state, encoding the evolution that causally relates $S$ and $R$, is then given by \cite{Leifer13}
\begin{align}
	\mathcal{P}_{R|S}^{\Lambda} = \sum_{m, n} \ketbra{n_{S}}{m_{S}}  \otimes \Lambda \left(\ketbra{m_{S^\prime}}{n_{S^\prime}}\right),
\end{align}
where $\{\ket{m_{S}} \}$ and $\{\ket{m_{S^\prime}} \}$ are the complete set of orthonormal bases spanning $\mathcal{H}_S$ and $\mathcal{H}_{S^\prime}$ respectively. Here $\mathcal{H}_{S^\prime}$ is a copy of $\mathcal{H}_S$ and $\Lambda: \mathcal{L}(\mathcal{H}_{S^\prime}) \mapsto \mathcal{L}(\mathcal{H}_R)$ as well.  A state transformation $\rho_R=\Lambda(\rho_S)$, where $\rho_S$ and $\rho_R$ are the density operators representing the states of $R$ and $S$ respectively, is equivalently expressed as $\rho_R = \tr_S \left[ \mathcal{P}_{R|S}^{\Lambda} \star \rho_S \right]$ with $X\star Y= Y^{\frac{1}{2}} X Y^{\frac{1}{2}}$. Now the quantum Bayes' rule \cite{Leifer13} can be cast as 
\begin{align}
\mathcal{P}_{S|R}^{\bar{\Lambda}} = \mathcal{P}_{R|S}^{\Lambda} \star (\rho_S  \otimes \rho_R^{-1}),	
\end{align} 
and it satisfies $\rho_S = \tr_S \left[ \mathcal{P}_{S|R}^{\bar{\Lambda}} \star \rho_R \right]$. Note,  $\mathcal{P}_{S|R}^{\bar{\Lambda}}$ is the causal conditional state corresponding to Petz recovery channel \cite{Petz88, Barnum02, Crooks08} or the inverse process 
\begin{align}
\bar{\Lambda}(\cdot):=\rho_S^{\frac{1}{2}} \Lambda^\dag \left(\rho_R^{-\frac{1}{2}} \ (\cdot) \ \rho_R^{-\frac{1}{2}} \right) \rho_S^{\frac{1}{2}},
\end{align} 
where $\Lambda^\dag$ is the trace dual of $\Lambda$, satisfying the relation $\tr [Y \ \Lambda(X)]=\tr[\Lambda^\dag(Y) \ X]$ for all operators $X$ and $Y$. 

The causal conditional state $\mathcal{P}_{S|R}^{\Lambda}$ corresponding to the inverse process depends on the reference prior $\rho_S$. However, while making inferences, this prior is often unknown. Then, there are two possible choices. One choice is to consider a known steady state $\rho_S=\gamma_S$ as prior, satisfying $\Lambda(\gamma_S)=\gamma_S$. Another is the uniform prior $\rho_S=\frac{\mathbb{I}}{d}$. The latter is obviously the viable choice in 'inverting' a process when no prior information is available.  Interestingly, the inverse of a deterministic (or unitary) process is independent of the reference prior \cite{Aw21}. For any isometric (or unitary) evolution $U: \mathcal{H}_S \mapsto \mathcal{H}_R$, where $U \ket{m_S} = \ket{m_R}$ for a complete set of orthonormal bases $\{\ket{m_S}\}$, the causal conditional states assume simpler forms
\begin{align}
&\mathcal{P}_{R|S}^{U}= \sum_{m, n} \ketbra{n_{S}}{m_{S}}  \otimes {U} \ketbra{m_{S^\prime}}{n_{S^\prime}}U^{\dag}, \\
& \mathcal{P}_{S|R}^{U^{\dag}}= \sum_{m, n} U^{\dag} \ketbra{n_{R^\prime}}{m_{R^\prime}}U  \otimes \ketbra{m_R}{n_R}, 
\end{align}  
where $R^\prime$ is the second copy of $R$ and $\bar{U}=U^\dag$. Note, $\mathcal{P}_{S|R}^{U^{\dag}}$ represents the evolution $U^{\dag}: \mathcal{H}_R \mapsto \mathcal{H}_S$, and $\mathcal{P}_{S|R}^{U^{\dag}}=\mathcal{P}_{R|S}^{U}$.

We can now establish deterministic causal relations between quantum cause and effect. For a general evolution $\Lambda: \mathcal{L}(\mathcal{H}_S) \mapsto \mathcal{L}(\mathcal{H}_R)$, the causal conditional states can be found. Here for the inverse process, we shall exploit uniform prior (or a steady state, whenever it is known). Say, after the process, one observes an effect by selectively measuring $R$ to find $\sigma_R$ and wants to infer the cause corresponding to it or vice versa. Then the conditions to draw deterministic causal relation are given in the following definition.

\begin{defi}[Deterministic quantum causal relation]\label{prop:GenCausal}
A cause $\tau_S$ deterministically predicts the effect $\tau_R$ due to the evolution $\Lambda$ if
\begin{align}
\tau_R=\tr_S \left[\mathcal{P}_{R|S}^\Lambda \star \tau_S \right]. \label{eq:GenCtoE}	
\end{align} 
In reverse, an observed effect $\sigma_R$ after the evolution by $\Lambda$ infers the cause $\sigma_S$ with certainty if
\begin{align}
	\sigma_S = \tr_R \left[\mathcal{P}_{S|R}^{\bar{\Lambda}} \star \sigma_R \right]. \label{eq:GenEtoC}
\end{align}
\end{defi}
Consider the earlier isometry $U$ leading to a (pure) state transformation $\ket{\psi_S} \mapsto \ket{\psi_R}$. An effect $\ket{\phi_R}=\sum_m a_{m} \ket{m_R}$ observed in the final state $\ket{\psi_R}$, upon a projective measurement using $\ketbra{\phi_R}{\phi_R}$, has one-to-one causal correspondence with the cause $\ket{\phi_S}=\sum_m b_{m} \ket{m_S}$ if they respect the relations
\begin{align}
&\ket{\phi_R} = U \ket{\phi_S}, \label{eq:tCtoE}\\
& \ket{\phi_S} = U^\dag \ket{\phi_R}, \label{eq:tEtoC}
\end{align}
which are equivalently the conditions~\eqref{eq:GenCtoE} and  \eqref{eq:GenEtoC} in Definition~\ref{prop:GenCausal}. Consequently, $\ket{\phi_S}$ and $\ket{\phi_R}$ have deterministic causal correspondence if $a_m=b_m, \ \forall m$.  

Now we turn to a situation where two quantum systems are evolved with a known global (i.e., non-local or entangling) evolution, and causes belonging to one system induce effects in the other. In particular, we focus on quantum measurement processes involving a system ($S$) and an apparatus ($A$), in which observations in the latter are used to infer about the former. Consider an evolution by a CPTP map $\Lambda_M: \mathcal{L}(\mathcal{H}_S\otimes \mathcal{H}_A) \to \mathcal{L}(\mathcal{H}_R \otimes \mathcal{H}_B)$. Without loss of generality, we assume $\mathcal{H}_R$ and $\mathcal{H}_B$ are the copies of $\mathcal{H}_S$ and $\mathcal{H}_A$, respectively. The composite $SA$ is initially in an uncorrelated state, say $\rho_{SA}=\rho_S \otimes \proj{0_A}$, where $\ket{0_A}$ is a known state of $A$.  The composite may become strongly correlated after the global evolution by $\Lambda_M$. Because of that, a local measurement on $B$ may induce a change in $R$. Say, a local measurement on $B$ results in an observation of a state $\sigma_B$, where the overall updated state is $\sigma_{RB}$ with $\sigma_B=\tr_R [\sigma_{RB}]$. Then, the effect $\sigma_B$ in $B$ establishes one-to-one causal correspondence with a cause $\sigma_S$ in $S$ if they satisfy the conditions involving the causal conditional states $\mathcal{P}^{\Lambda_M}_{RB|SA}$ and $\mathcal{P}^{\bar{\Lambda}_M}_{SA|RB}$ given in the definition below.

\begin{defi}[Deterministic quantum causal relation between cause and effect belonging to different systems]\label{prop:MeasCausal}
A cause $\tau_S$ in $S$ deterministically predicts the effect $\tau_B=\tr_R[\tau_{RB}]$ in $B$ via the evolution by $\Lambda_M$ if
\begin{align}
\tau_{RB} = \tr_{SA}\left[ \mathcal{P}^{\Lambda_M}_{RB|SA} \star \tau_S \otimes \proj{0_A} \right], \label{eq:CtooE}
\end{align}
where $\tau_S \equiv \tr_B[\tau_{RB}]$. In reverse, an observed effect $\sigma_B=\tr_R[\sigma_{RB}]$ in $B$ deterministically infers the cause $\sigma_S$ in $S$ if
\begin{align}
	\sigma_S \otimes \proj{0_A} = \tr_{RB} \left[\mathcal{P}^{\bar{\Lambda}_M}_{SA|RB} \star \sigma_{RB}  \right], \label{eq:EtooC}
\end{align}
where $\sigma_S \equiv \tr_B[\sigma_{RB}]$.
\end{defi}   

Traditionally, an ideal quantum measurement process involves coherent copying (i.e., generalized C-NOT) operation $U_{mes}: \mathcal{H}_S \otimes \mathcal{H}_A \mapsto \mathcal{H}_R \otimes \mathcal{H}_B$. Then, an arbitrary state $\ket{\psi_S}=\sum_i c_i \ket{i_S}$ of $S$ leads to
\begin{align}
\ket{\psi_S} \ket{0_A}\xrightarrow{U_{mes}} \sum_i c_i \ket{i_R} \ket{i_B}, \label{eq:IdMeas}
\end{align}
where $\ket{k}\ket{l}=\ket{k} \otimes \ket{l}$. Unlike in classical cases, in this quantum evolution, $S$ and $A$ both may causally influence $R$ and $B$  \cite{Allen17, Barrett19}. Say, one observes an effect $\ket{\phi_B}=\sum_{i}k_i \ket{i_B}$ in $B$ after implementing the (rank-$1$) projector $\ketbra{\phi_B}{\phi_B}$ on $B$ and, consequently the updated $RB$ state becomes $\ket{\phi_{RB}^\prime}=\frac{1}{N}\sum_i c_i k_i^* \ket{i_R} \ket{\phi_B}=\ket{\phi_R}\ket{\phi_B}$, where $N=(\sum_i |c_i k_i^*|^2)^{1/2}$. Note the collapse induced in $R$ due to the observation on $B$. Now, a cause $\ket{\phi_S}$ belonging to $S$ has one-to-one causal correspondence with the effect $\ket{\phi_B}$ in $B$ if
\begin{align}
&U_{mes}\ket{\phi_S} \ket{0_A} =\ket{\phi_R} \ket{\phi_B}, \label{eq:sCtoE} \\ 
&U_{mes}^\dag \ket{\phi_R} \ket{\phi_B} =\ket{\phi_S} \ket{0_A},	\label{eq:sEtoC}
\end{align}
which are exactly the conditions \eqref{eq:CtooE} and \eqref{eq:EtooC} in Definition~\ref{prop:MeasCausal}. Here $\ket{\phi_S}=\ket{\phi_R}$. Hence, only the effects  $\{ \ket{i_B}\}$ establishes one-to-one correspondence with the causes $\{ \ket{i_S}\}$ respectively. Any other effect in $B$ will not establish such a correspondence with a cause in $S$ and vice versa. 

In general, the situation for unitary evolution is simpler than the non-unitary ones. This is due to the fact that unitary evolutions preserve all information and are invertible (without a need for a reference prior), and thus it is sufficient to check conditions \eqref{eq:tCtoE} and \eqref{eq:sCtoE} for deterministic causal predictions and  conditions  \eqref{eq:tEtoC} and \eqref{eq:sEtoC} for deterministic causal inferences. In the rest of the article, we restrict ourselves to the cases that involve unitary evolution and measurements using rank-1 projectors.

\section{ Classical vs quantum causal relations \label{sec:CCRvsQCR}}
In this section, we shall study the situations that will help reveal the inadequacy of CCR. In particular, we consider two examples where causal inferences based on CCR differ from those based on QCR.  \\

\begin{figure}[t]
	\includegraphics[width=0.95\columnwidth]{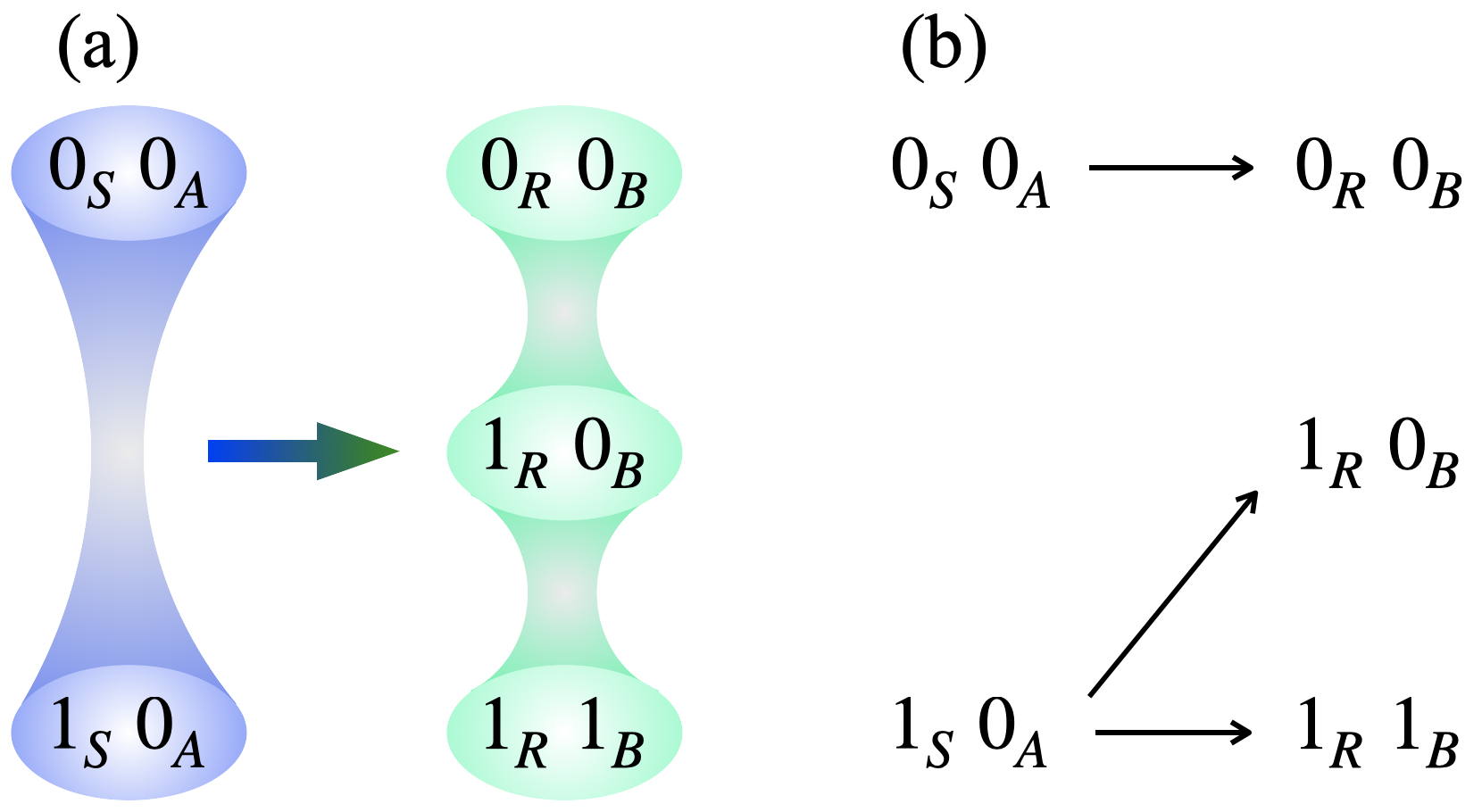}
	\caption{{\bf Example-1.} (a) The quantum process~\eqref{eq:USupEff}-\eqref{eq:EtoC} allows superposition among orthogonal causes and, as a result, can result in an effect that is in a superposition between orthogonal effects.  (b) The figure displays the classical process analogous to the quantum one in Eq.~\eqref{eq:USupEff}. Here, the classical bit $0_S0_A$ deterministically evolves to $0_R0_B$, as the conditional probability is $P(0_R0_B|0_S0_A)=1$. But, bits $1_S0_A$ results in incoherent mixtures of $1_R0_B$ and $1_R1_B$, with conditional probabilities $P(1_R0_B|1_S0_A)=\frac{1}{2}$ and $P(1_R1_B|1_S0_A)=\frac{1}{2}$. See text for more details. \label{fig:Example1}}
\end{figure}

{\it Example-1:}  This example involves an initially uncorrelated system-apparatus composite and a global unitary evolution (see Figure~\ref{fig:Example1}). Say both system ($S$) and apparatus ($A$) are qubits, and they are evolved together with an isometry $V: \mathcal{H}_S \otimes \mathcal{H}_A \mapsto \mathcal{H}_R \otimes \mathcal{H}_B$, given by
\begin{align}
 \ket{0_S} \ket{0_A} \xrightarrow{V} \ket{0_R} \ket{0_B}, \ \ \ \ket{1_S} \ket{0_A} \xrightarrow{V} \ket{1_R} \ket{+_B}, \label{eq:USupEff}
\end{align}
where $\{\ket{0_X}, \ \ket{1_X}\}$ are the orthonornal bases of $\mathcal{H}_X$ and $\ket{\pm_X}=1/\sqrt{2}(\ket{0_X} \pm \ket{1_X})$. The initial state is $\ket{\phi_S} \ket{0_A}$ with $\ket{\phi_S}=\sqrt{1-r}\ket{0_S} + \sqrt{r} \ket{1_S}$ for $0 < r < 1$. After the evolution by $V$, the final  state $\ket{\phi_{RB}}=V \ket{\phi_S} \ket{0_A}$ becomes
\begin{align}
	\ket{\phi_{RB}} & = \sqrt{1-r} \ \ket{0_R} \ket{0_B} + \sqrt{r} \ \ket{1_R} \ket{+_B},  \label{eq:CtoE} \\ 
	& = \left(\sqrt{1-r} \ \ket{0_R} + \sqrt{r/2} \ \ket{1_R} \right) \ket{0_B} + \sqrt{r/2} \ \ket{1_R} \ket{1_B}.  \label{eq:EtoC}
\end{align}
The evolution by $V$ does not establish deterministic causal relations for arbitrary causes and effects. Some causes like $\ket{0_S}$ and $\ket{1_S}$ can predict the effects $\ket{0_A}$ and $\ket{+_A}$ with certainty. However, the observations of $\ket{0_A}$ and $\ket{+_A}$ cannot be used to deterministically infer the causes $\ket{0_S}$ and $\ket{1_S}$ respectively.

To highlight how CCR differs from QCR, let us analyze the observation of the effect $\ket{1_B}$ in $B$ and infer its cause in $S$. Due to the presence of entanglement in the state \eqref{eq:EtoC}, a local observation of $\ket{1_B}$ induces a collapse in $R$ to the state $\ket{1_R}$. Because of that, we need to consider the causal correspondence between global causes and effects. Thus, the task is now to find the cause in $SA$ for the observed effect $\ket{1_R}\ket{1_B}$ in $RB$. To infer the cause using CCR, we consider a classical stochastic process analogous to the evolution \eqref{eq:USupEff}, see Figure~\ref{fig:Example1}(b). There, the (classical) conditional probability for $\ket{1_S}\ket{0_A} \mapsto \ket{1_R}\ket{1_B}$ is $P(1_R1_B|1_S0_A)=\frac{1}{2}$, and the probabilities of finding $\ket{1_S}\ket{0_A}$ and $\ket{1_R} \ket{1_B}$ before and after the global evolution are $P(1_S0_A)=r$ and $P(1_R1_B)=\frac{r}{2}$ respectively. Then the conditional probability $P(1_S0_A|1_R1_B)$ representing the inverted evolution can be derived using classical Bayes' rule, and that is
\begin{align}
	P(1_S0_A|1_R1_B)=P(1_R1_B|1_S0_A) \ P(1_S0_A)/P(1_R1_B)=1.
\end{align}
Following CCR based on classical Bayes' rule, this unit conditional probability implies that the effect $\ket{1_R}\ket{1_B}$ can deterministically infer the cause $\ket{1_S}\ket{0_A}$. This, in turn, means that $\ket{1_S}$ in $S$ is the cause for the effect $\ket{1_B}$ observed in $B$. However, this cannot be true because it does not satisfy the condition~\eqref{eq:EtooC} or \eqref{eq:sEtoC} for deterministic quantum causal inference based on quantum Bayes' rule, as
\begin{align}
	\tr_{RB}\left[ \mathcal{P}_{SA|RB}^{V^\dag} \star \ketbra{1_R}{1_R} \otimes \ketbra{1_B}{1_B}\right]\neq \ketbra{1_S}{1_S} \otimes \ketbra{0_A}{0_A}, \nonumber
\end{align} 
or $V^\dag \ket{1_R} \ket{1_B} = \ket{1_S}\ket{-_A} \neq \ket{1_S} \ket{0_A}$. Thus, according to QCR, the effect $\ket{1_B}$ cannot deterministically infer the cause $\ket{1_S}$. It is at most be claimed that the cause $\ket{1}_S$ may result in the effect $\ket{1}_B$ with probability $\frac{1}{2}$. This is in direct contradiction with the inference made using CCR.  \\

\begin{figure}[t]
	\includegraphics[width=0.95\columnwidth]{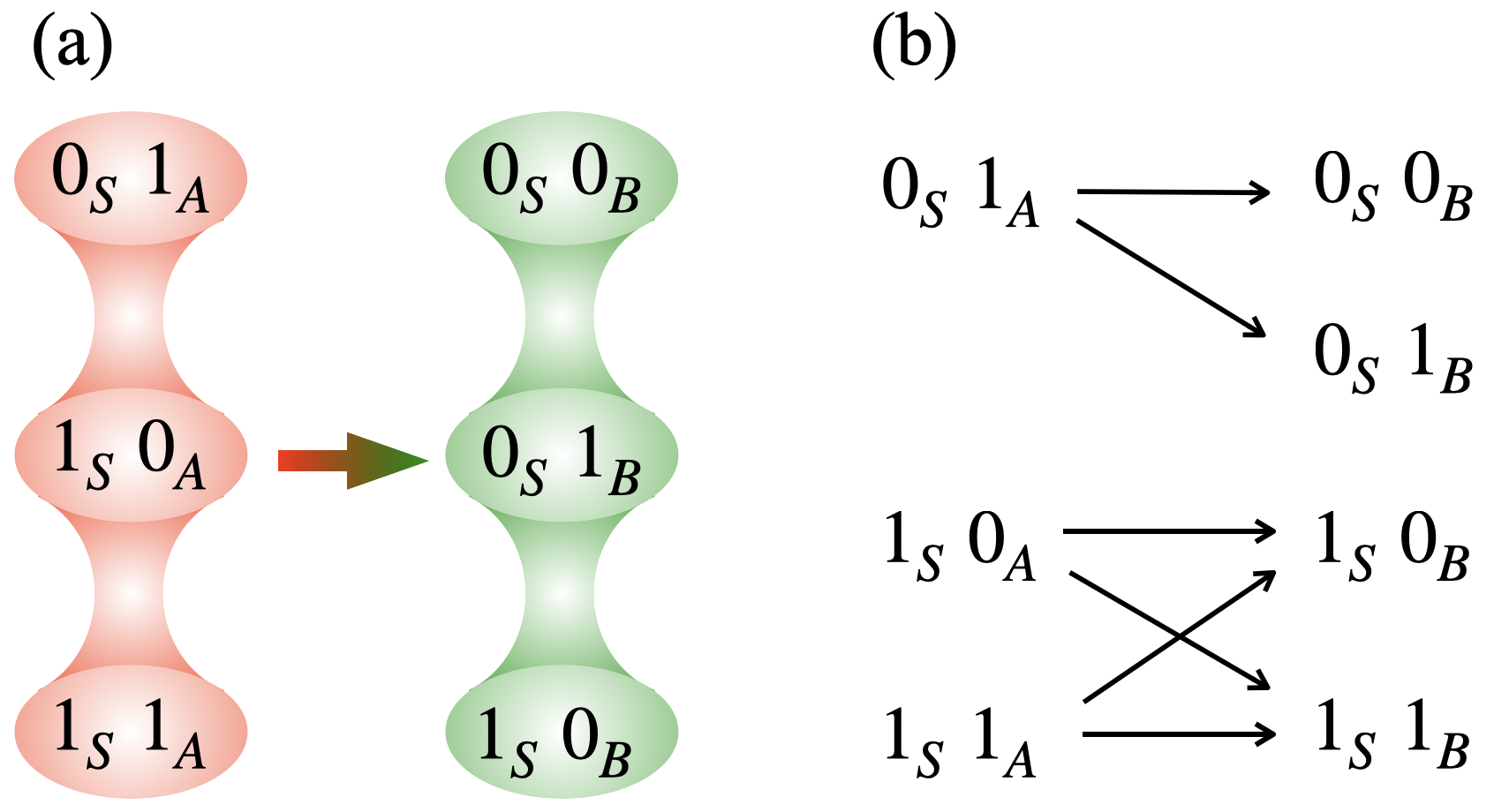}
	\caption{{\bf Example-2.} 	\label{fig:Example2} (a) The quantum evolution is depicted. It allows superposition causes in $SA$ to result in a superposition of effects in $SB$. (b) The figure represents analogous classical stochastic evolution where classical bits  $SA$ act as the cause to induce incoherent mixtures of effects in $SB$, with all classical conditional probabilities $P(0_S0_B|0_S1_A)$, $P(0_S1_B|0_S1_A)$, $P(1_S0_B|1_S0_A)$, $P(1_S1_B|1_S0_A)$, $P(1_S0_B|1_S1_A)$, and $P(1_S1_B|1_S1_A)$ equal to $\frac{1}{2}$. See text for more details. }
\end{figure}

{\it Example 2} -- This example assumes a situation where the initial state of the system ($S$) and apparatus ($A$) composite is entangled and evolves via local unitary operations (see Figure~\ref{fig:Example2}). Say a two-qubit composite $SA$ is in an initially entangled state
\begin{align}
	\ket{\psi_{SA}}=\frac{1}{\sqrt{3}}\left(i \ket{0_S}\ket{1_A} + i \ket{1_S}\ket{0_A} + \ket{1_S}\ket{1_A} \right), \label{eq:exmp2_inState}
\end{align}
where $\{\ket{0_X}, \ \ket{1_X}\}$ are the orthonornal bases of $\mathcal{H}_X$. The qubit $A$ undergoes an evolution by the isometry $U_A: \mathcal{H}_A \mapsto \mathcal{H}_B$, with
\begin{align}
	\ket{0_A} \mapsto \frac{1}{\sqrt{2}} \left( \ket{0_B} + i \ket{1_B} \right), \ \  \ket{1_A} \mapsto \frac{1}{\sqrt{2}} \left(i \ket{0_B} + \ket{1_B} \right),
\end{align}
to result in the final state $\ket{\phi_{SB}}=\mathbb{I}_S \otimes U_A\ket{\psi_{SA}}$ given by
\begin{align}
\ket{\phi_{SB}}=\frac{1}{\sqrt{6}} \left(2i \ket{1_S}\ket{0_B} - \ket{0_S}\ket{0_B} + i \ket{0_S}\ket{1_B}\right). 	
\end{align}
Here the evolution implemented is local in nature and thus cannot establish causal relations between $S$ and $B$ in general. But, initial entanglement may result in a causal correspondence between $S$ and $B$. Thus the causal relation must involve the global causes in $SA$ and global effects in $SB$.

In order to highlight how inferences based on classical and quantum causal relations differ, let us now find out the cause (in $SA$) corresponding to the effect $\ket{0_S}\ket{1_B}$ observed in the final state $\ket{\phi_{SB}}$. For that, an analogous classical stochastic process is depicted in Figure~\ref{fig:Example2}(b). With the classical conditional probability $P(0_S1_B|0_S1_A)=\frac{1}{2}$ for the transition $\ket{0_S}\ket{1_A} \mapsto \ket{0_S}\ket{1_B}$ and probabilities of finding  $\ket{0_S}\ket{1_A}$ and $\ket{0_S}\ket{1_B}$ in the initial and final states $P(0_S1_A)=\frac{1}{3}$ and $P(0_S1_B)=\frac{1}{6}$ respectively, the classical Bayes' rule leads to
\begin{align}
	P(0_S1_A| 0_S1_B)=P(0_S1_B|0_S1_A) \ P(0_S1_A)/P(0_S1_B)=1.
\end{align}
Thus, as per CCR, an observation of the effect $\ket{0_S}\ket{1_B}$ deterministically infers the cause  $\ket{0_S}\ket{1_A}$. This may be further argued by the facts that the observation of $\ket{1_B}$ happens together with $\ket{0_S}$ in $\ket{\phi_{SB}}$, and $\ket{0_S}$ in $\ket{\phi_{SB}}$ implies $\ket{0_S}$ in $\ket{\psi_{SA}}$ because in the transformation $\ket{\psi_{SA}} \mapsto \ket{\phi_{SB}}$ the qubit $S$ did not evolve. Altogether, the observation of the effect $\ket{1_B}$ in $\ket{\phi_{SB}}$ demands the cause $\ket{0_S}$ to be present in $\ket{\psi_{SA}}$. But, this inference cannot be true. According to quantum Bayes' rule
\begin{align}
	\tr_{SB}\left[ \mathcal{P}_{SA|SB}^{U^\dag_{A}} \star \ketbra{0_S}{0_S} \otimes \ketbra{1_B}{1_B}\right]\neq \ketbra{0_S}{0_S} \otimes \ketbra{1_A}{1_A}, \nonumber
\end{align} 
or $U^\dag_{A}\ket{0_S}\ket{1_B} = \frac{1}{\sqrt{2}}( \ket{0_S}\ket{1_A}- i \ket{0_S}\ket{0_A}) \neq \ket{0_S}\ket{1_A}$, and it does not satisfy the condition~\eqref{eq:GenEtoC} or \eqref{eq:tEtoC} for deterministic causal inference. One may, at most, claim that the cause $\ket{0_S}\ket{1_A}$ is responsible for the effect $\ket{0_S}\ket{1_B}$ with probability $\frac{1}{2}$. 

A similar problem also appears in case of prediction. For instance, consider the prediction of the effect $\ket{1_S}\ket{1_B}$ in the final state $\ket{\phi_{SB}}$. The probabilities of finding $\ket{1_S}\ket{0_A}$ and $\ket{1_S}\ket{1_A}$ in $\ket{\phi_{SA}}$ are $P(1_S0_A)=\frac{1}{3}$ and $P(1_S1_A)=\frac{1}{3}$ respectively. The (classical) transition probabilities for $\ket{1_S}\ket{0_A} \mapsto \ket{1_S}\ket{1_B}$ and $\ket{1_S}\ket{1_A} \mapsto \ket{1_S}\ket{1_B}$ are $P(1_S1_B|1_S0_A)=\frac{1}{2}$ and $P(1_S1_B|1_S1_A)=\frac{1}{2}$ respectively. Then, according to CCR, the effect $\ket{1_S}\ket{1_B}$ should be observed with a probability
\begin{align*}
P(1_S1_B|1_S0_A) P(1_S0_A) + P(1_S1_B|1_S1_A) P(1_S1_A)=\frac{1}{3}.	
\end{align*}
However, according to QCR, the effect $\ket{1_S}\ket{1_B}$ can never be observed in the final state, which is indeed the case. 

Therefore, the inferences and predictions based on CCR differ from the ones based on QCR. The use of CCR, in fact, leads to various contradictions in quantum mechanics, and that can again be resolved with the help of QCR. Two such cases are considered in the following sections. 

\section{Frauchiger-Renner's paradox and QCR \label{sec:FRvsQCR}}
We now consider a case based on the setup assumed in Frauchiger-Renner's paradox \cite{Frauchiger18}. The paradox is an extension of Wigner's friend paradox \cite{Wigner67} and shows that the predictions and inferences about a system by different observers and super-observers at different stages of evolution are inconsistent with each other. This, in turn, introduces a no-go theorem between various interpretations of quantum mechanics and claims that ``quantum mechanics cannot consistently describe the use of itself''. 

Here we re-visit Frauchiger-Renner's Paradox. In this setup, the initial states of the systems involved and the measurement (unitary) evolutions are known a priory. Without bringing in technicalities, the paradox is demonstrated using the steps below. \\

\noindent 
(F0) At first, a qubit $R$ is prepared in the initial state
\begin{align}
	\ket{\psi_R}=\frac{1}{\sqrt{3}}\ket{h_R} + \sqrt{\frac{2}{3}} \ket{t_R}. \label{eq:s0}
\end{align}

\noindent 
(F1) Then, $R$ is attached with a spin-$\frac{1}{2}$ system $S$ and evolved using the isometry $V_1$ given by
\begin{align}
	\ket{h_R} \mapsto \ket{\bar{h}_{\bar{L}}} \ket{\downarrow_S}, \ \ \ \ket{t_R} \mapsto \ket{\bar{t}_{\bar{L}}} \ket{\rightarrow_S},
\end{align}
where $\ket{\rightleftarrows_S}=\frac{1}{\sqrt{2}}(\ket{\downarrow_S} \pm \ket{\uparrow_S})$. The isometry $V_1$ updates the initial state $\ket{\psi_R}$ to 
\begin{align}
	\ket{\psi_{\bar{L} S}}&=\frac{1}{\sqrt{3}}\left(\ket{\bar{h}_{\bar{L}}} \ket{\downarrow_S} + \ket{\bar{t}_{\bar{L}}} \ket{\downarrow_S} + \ket{\bar{t}_{\bar{L}}} \ket{\uparrow_S} \right), \label{eq:s1.1} \\
	&= \sqrt{\frac{2}{3}}\ket{\bar{f}_{\bar{L}}} \ket{\downarrow_S} + \frac{1}{\sqrt{6}}\ket{\bar{f}_{\bar{L}}} \ket{\uparrow_S} - \frac{1}{\sqrt{6}}\ket{\bar{o}_{\bar{L}}} \ket{\uparrow_S}, \label{eq:s1.2}  
\end{align}
where in the second step we have used $\ket{\bar{t}_{\bar{L}}}=\frac{1}{\sqrt{2}}(\ket{\bar{f}_{\bar{L}}}-\ket{\bar{o}_{\bar{L}}})$ and $\ket{\bar{h}_{\bar{L}}}=\frac{1}{\sqrt{2}}(\ket{\bar{f}_{\bar{L}}}+\ket{\bar{o}_{\bar{L}}})$. 

\noindent 
(F2) Now another isometry $V_2$ is applied on $S$, given by
\begin{align}
	\ket{\downarrow_S} \mapsto \frac{1}{\sqrt{2}}(\ket{f_L}+\ket{o_L}), \ \ \ \ket{\uparrow}_S \mapsto \frac{1}{\sqrt{2}}(\ket{f_L}-\ket{o_L}),
\end{align}
and as consequence, the state $\ket{\psi_{\bar{L} S}}$ modifies to
\begin{align}
	\ket{\psi_{\bar{L} L}} = \frac{1}{\sqrt{12}}\left( \ket{\bar{o}_{\bar{L}}}(\ket{o_{L}} - \ket{f_{L}}) + \ket{\bar{f}_{\bar{L}}}(\ket{o_{L}} + 3 \ket{f_{L}}) \right). \label{eq:s2}
\end{align}
The contradiction leading to the paradox can be understood by noting inconsistencies in the chain of arguments based on classical Bayes' theorem below, similar to the ones considered in \cite{Frauchiger18}. (A1) An observation of $\ket{\bar{o}_{\bar{L}}}$ in the state $\ket{\psi_{\bar{L}S}}$ ensures the observation of $\ket{\uparrow_S}$ (see Eq.~\eqref{eq:s1.2}). (A2) Again from $\ket{\psi_{\bar{L}S}}$, it is guaranteed that the observation of $\ket{\uparrow_S}$ always occurs together with the observation of $\ket{\bar{t}_{\bar{L}}}$ (see Eq.~\eqref{eq:s1.1}). (A3) From the action of the isometry $V_1$ in step (F1), it is ``inferred'' that the observation of $\ket{\bar{t}_{\bar{L}}}$ has the underlying cause $\ket{t_R}$. (A4) The cause $\ket{t_R}$ guarantees that the state of $S$ is $\ket{\rightarrow_S}$ after the evolution by $V_1$. (A5) With the evolution by $V_2$, the cause $\ket{\rightarrow_S}$ in $S$ leads to the effect $\ket{f_{L}}$ in $L$. (A6) As seen from the state $\ket{\psi_{\bar{L} L}}$ (see Eq.~\eqref{eq:s2}) in step (F2), the joint state $\ket{\bar{o}_{\bar{L}}} \ket{o_{L}}$ is observed with the probability $\frac{1}{12}$.

Using arguments (A1)-(A3), it is ``inferred'' that the cause of the effect $\ket{\bar{o}_{\bar{L}}}$ in $\bar{L}$ is $\ket{t_R}$ and, following the arguments (A4)-(A5), this cause predicts the effect $\ket{f_{L}}$ in $L$. Therefore, each observation of $\ket{\bar{o}_{\bar{L}}}$ is associated with the observation of $\ket{f_{L}}$. This is equivalent to say that the joint state $\ket{\bar{o}_{\bar{L}}}\ket{o_{L}}$ should never be observed. However, the argument (A6) claims that $\ket{\bar{o}_{\bar{L}}}\ket{o_{L}}$ will be observed with a non-zero probability. This leads to a contradiction and the paradox. 

The root of this apparent inconsistency lies in the ignorance of the role of quantum evolution, measurement induced collapse while making inferences and predictions, and that quantum causes (effects) may coherently superpose to represent another cause (effects). Let us start by re-analyzing step (F1). The isometry $V_1$ can be implemented in two stages. First, an isometry $V_1^{(1)}$ that maps $R$ as $\ket{h_R} \mapsto \ket{\bar{h}_{\bar{L}}}$ and $\ket{t_R} \mapsto \ket{\bar{t}_{\bar{L}}}$. Then the system $S$ in a state $\ket{\downarrow_S}$ is clubbed with the $\bar{L}$ and a global unitary $V_1^{(2)}$ is applied on $\bar{L}S$, where
\begin{align}
	\ket{\bar{h}_{\bar{L}}}\ket{\downarrow_S} \xrightarrow{V_1^{(2)}} \ket{\bar{h}_{\bar{L}}}\ket{\downarrow_S}, \ \ \ \ket{\bar{t}_{\bar{L}}}\ket{\downarrow_S} \xrightarrow{V_1^{(2)}} \ket{\bar{t}_{\bar{L}}}\ket{\rightarrow_S}.
\end{align}
The overall isometry becomes $V_1=V_1^{(2)}\circ V_1^{(1)}$ as required. The unitary $V_1^{(2)}$ has properties similar to the unitary $V$ considered in Example-1 (see Eq.~\eqref{eq:USupEff}). It is true that the state $\ket{\uparrow_S}$ of $S$ always appears with the state $\ket{\bar{t}_{\bar{L}}}$ of $\bar{L}$ (argument (A2)). The corresponding probability of observing $\ket{\bar{t}_{\bar{L}}}\ket{\uparrow_S}$ in $\ket{\psi_{\bar{L} S}}$ is $P(\bar{t}_{\bar{L}}\uparrow_S)=\frac{1}{3}$. The probability of finding $\ket{\bar{t}_{\bar{L}}}\ket{\downarrow_S}$, before the application of $V_1^{(2)}$, i.e., in the state $V_1^{(1)}\ket{\psi_R}\ket{\downarrow_S}$, is $P(\bar{t}_{\bar{L}}\downarrow_S)=\frac{2}{3}$. Now with the classical conditional probability $P(\bar{t}_{\bar{L}}\uparrow_S|\bar{t}_{\bar{L}}\downarrow_S)=\frac{1}{2}$ for the transition $\ket{\bar{t}_{\bar{L}}}\ket{\downarrow_S} \to \ket{\bar{t}_{\bar{L}}}\ket{\uparrow_S}$ and using classical Bayes' rule, one finds
\begin{align}
P(\bar{t}_{\bar{L}}\downarrow_S|\bar{t}_{\bar{L}}\uparrow_S)=P(\bar{t}_{\bar{L}}\uparrow_S|\bar{t}_{\bar{L}}\downarrow_S) P(\bar{t}_{\bar{L}}\downarrow_S)/P(\bar{t}_{\bar{L}}\uparrow_S)=1.	
\end{align} 
With the (inverted) conditional probability equals to one, as per CCR, the observation of the effect $\ket{\bar{t}_{\bar{L}}}\ket{\uparrow_S}$ deterministically infers the cause $\ket{\bar{t}_{\bar{L}}}\ket{\downarrow_S}$. Similarly, $\ket{\bar{t}_{\bar{L}}}$ in $\bar{L}$ is attributed to the cause $\ket{t_R}$ in $R$, due to the (inverted) evolution $V^{(1)\dag}_1$. This is exactly the basis for the argument (A3). 

However, as we have discussed in Example-1, the argument (A3) cannot be true. Because it does not respect the conditions for a deterministic quantum causal relation. That is, according to quantum Bayes' rule,
\begin{align}
	\tr_{\bar{L}S}\left[ \mathcal{P}_{\bar{L}S|\bar{L}S}^{V_1^{(2)\dag}} \star \ketbra{\bar{t}_{\bar{L}}}{\bar{t}_{\bar{L}}} \otimes \ketbra{\uparrow_S}{\uparrow_S}\right]\neq \ketbra{\bar{t}_{\bar{L}}}{\bar{t}_{\bar{L}}} \otimes \ketbra{\downarrow_S}{\downarrow_S}, \nonumber
\end{align} 
or equivalently, $V_1^{(2)\dag}\ket{\bar{t}_{\bar{L}}}\ket{\uparrow_S}=
\ket{\bar{t}_{\bar{L}}}\ket{\leftarrow_S} \neq \ket{\bar{t}_{\bar{L}}}  \ket{\downarrow_S}$. Thus, the inference drawn from the arguments (A2) and (A3) is incomplete. It can at most be said that the observed global effect $\ket{\bar{t}_{\bar{L}}} \ket{\uparrow_S}$ is the result of a global cause $\ket{\bar{t}_{\bar{L}}}\ket{\leftarrow_S}$, where the cause $\ket{\bar{t}_{\bar{L}}} \ket{\downarrow_S}$ is present with probability $\frac{1}{2}$. Given this global cause, the isometry $V_2$ in step (F2) guarantees that the effect $\ket{\bar{o}_{\bar{L}}} \ket{o_{L}}$ is observed with a non-zero probability, since
\begin{align}
	\ket{\bar{t}_{\bar{L}}}\ket{\leftarrow_S} \mapsto \ket{\bar{t}_{\bar{L}}}\ket{o_L}=\frac{1}{\sqrt{2}}(\ket{o_L}\ket{\bar{f}_{\bar{L}}} -\ket{o_L} \ket{\bar{o}_{\bar{L}}}).
\end{align}
Clearly, the conclusion drawn from the arguments (A1)-(A5) earlier is untrue. Furthermore, we can easily see that the $\ket{\psi_{\bar{L} S}}$ represents a superposition of other global causes, where  $\ket{\bar{t}_{\bar{L}}}\ket{\uparrow_S}$ being one of them. All these causes together in superposition, i.e., the cause $\ket{\psi_{\bar{L} S}}$, lead to a the observation of the effect $\ket{\bar{o}_{\bar{L}}}\ket{o_L}$ with the probability $\frac{1}{12}$ upon application of $V_2 \circ V_1^{(2)\dag}$ or QCR. This agrees with the argument (A6). Therefore, there is no contradiction (or paradox) once one uses QCR while making deterministic predictions or inferences.

\section{Hardy's setup and QCR \label{sec:HvsQCR}}
Below, we reconsider Hardy's exposition of nonlocality in a bipartite system \cite{Hardy93} where his paradox \cite{Hardy92} becomes a special case of the nonlocal feature. We show that there is still a contradiction even after assuming that quantum mechanics is a non-local theory. And this is exclusively due to the use of CCR. Consider a bipartite system composed of two qubits $M$ and $N$ in a state
\begin{align}
	\ket{\psi_{MN}}=\alpha \ket{0_M}\ket{0_N} + \beta \ket{1_M} \ket{1_N},
\end{align}
where $\alpha, \ \beta \in \mathbb{R}$, $|\alpha| \neq |\beta|$, and satisfy $|\alpha|^2 + |\beta|^2=1$. Here $\{\ket{0_X}, \ \ket{1_X}\}$ is the orthonormal basis set spanning the Hilbert space $\mathcal{H}_X$ of the qubit $X$. The state can be re-expressed in a new orthonormal basis set
\begin{align}
	\ket{\psi_{MN}}=\sqrt{\alpha \beta} \ \left(\ket{u_M}\ket{v_N} + \ket{v_M}\ket{u_N}\right) + (|\alpha| - |\beta|) \ \ket{v_M}\ket{v_N}	
\end{align}
after dropping the overall factor $-1$, where $\ket{0_X}=B \ket{u_X} + iA^* \ket{v_X}$ and $\ket{1_X}= i A\ket{u_X} + B^* \ket{v_X}$ with $A=\sqrt{\alpha}/\sqrt{|\alpha| + |\beta|}$ and $B=i\sqrt{\beta}/\sqrt{|\alpha| + |\beta|}$. Now the state is evolved with two local unitaries (isometries) $U_M: \mathcal{H}_M \mapsto \mathcal{H}_R$ and $U_N: \mathcal{H}_N \mapsto \mathcal{H}_S$ given by
\begin{align*}
	\ket{u_{M/N}} \mapsto a^* \ket{c_{R/S}} - b \ket{d_{R/S}}, \ \ \ket{v_{M/N}} \mapsto b^* \ket{c_{R/S}} + a \ket{d_{R/S}},
\end{align*}
where $a=\sqrt{\alpha \beta}/\sqrt{1-|\alpha \beta|}$ and $b=(|\alpha| - |\beta|)/\sqrt{1-|\alpha \beta|}$. Two sequences of unitaries are applied on the initial state leading to the same end state, as
\begin{align}
	&\ket{\psi_{MN}} \xrightarrow{{U_M} \otimes \mathbb{I}} \ket{\psi_{RN}} \xrightarrow{\mathbb{I} \otimes U_N} \ket{\psi_{RS}}, \nonumber \\
	\mbox{and} \ \ 	&\ket{\psi_{MN}} \xrightarrow{\mathbb{I} \otimes U_N} \ket{\psi_{MS}} \xrightarrow{U_M \otimes \mathbb{I}} \ket{\psi_{RS}}. \nonumber
\end{align}
Then, the updated states are
\begin{align*}
	&\ket{\psi_{RN}}=n\left[ \ket{c_R}\left(a \ket{u_N} + b \ket{v_N}\right) - a^2 \left(a^* \ket{c_R} - b \ket{d_R}\right) \ket{u_N} \right], \\
	&\ket{\psi_{MS}}= n\left[ \left(a \ket{u_M} + b\ket{v_M}\right)\ket{c_S} - a^2 \ket{u_M} \left(a^* \ket{c_S} -b \ket{d_S}\right) \right], \\
	& \ket{\psi_{RS}}=n \left[ \ket{c_R} \ket{c_S} - a^2\left(a^* \ket{c_R} - b\ket{d_R}\right)\left(a^* \ket{c_S} - b \ket{d_S}\right)  \right],
\end{align*}
where the normalization constant $n=(1-|\alpha \beta|)/(|\alpha| - |\beta|)$. Below we expose an irreconcilable contradiction in this general setting, even after accepting quantum mechanics as a non-local theory, in terms of the statements based on classical Bayes' rule.

For instance, (H0) we never observe $\ket{u_M}\ket{u_N}$ in the state $\ket{\psi_{MN}}$ because former is absent in the latter. (H1) From $\ket{\psi_{RS}}$ is it evident that the effect $\ket{d_R}\ket{d_S}$ is found upon a ``global'' observation with a probability given by $|na^2b^2| \neq 0$. (H2) Each observation of $\ket{d_R}\ket{d_S}$ in $\ket{\psi_{RS}}$ deterministically infers the cause $\ket{d_R}\ket{u_N}$ in $\ket{\psi_{RN}}$. (H3) Again, the observation of $\ket{u_N}$ in $\ket{\psi_{RN}}$ implies the presence of $\ket{u_N}$ in $\psi_{MN}$ as the evolution $\ket{\psi_{MN}} \mapsto \ket{\psi_{RN}}$ only locally updates the qubit $M$ without altering $N$. (H4) Each observation of $\ket{d_R}\ket{d_S}$ in $\ket{\psi_{RS}}$ deterministically infers the cause $\ket{u_M}\ket{d_S}$ in $\ket{\psi_{MS}}$. (H5) And, the observation of $\ket{u_M}$ in $\ket{\psi_{MS}}$ demands the presence of $\ket{u_M}$ in $\ket{\psi_{MN}}$, as the evolution $\ket{\psi_{MN}} \mapsto \ket{\psi_{MS}}$ only locally updates the qubit $N$. Now, the statements (H1)-(H5) together imply that the observation of the effect $\ket{d_R}\ket{d_S}$ in $\ket{\psi_{RS}}$ must has the cause $\ket{u_M}\ket{u_N}$ in $\ket{\psi_{MN}}$. However, this directly contradicts with the statement (H0).   

Now we re-investigate the conclusions drawn from the arguments above and their contradiction in light of quantum causal relation. It is worth noting that, contrary to Frauchiger-Renner's paradox, the local nature of the evolutions here does not necessarily establish a correspondence between the cause belonging to one system with the effect resulting in the other and vice versa. However, the presence of initial entanglement may establish some correspondence. Therefore, the causal analysis of the arguments requires simultaneously considering global cause and effect belonging to both systems.

What we demonstrate now is that the statements leading to the contradiction rely on CCR, and how the contradiction disappears once QCR is used for making inferences and predictions. Here the situation is similar to the case considered in Example-2. Let us reanalyze the statement (H2) and identify the cause in $RN$ corresponding to the effect $\ket{d_R}\ket{d_S}$ observed in $RS$ and, in particular, what CCR and QCR infer. The classical conditional probability $P(d_Rd_S|d_Ru_N)=|b|^2$ for the transition $\ket{d_R}\ket{u_N} \mapsto \ket{d_R}\ket{d_S}$. The probabilities of finding $\ket{d_R}\ket{u_N}$ in $\ket{\psi_{RN}}$ and $\ket{d_R}\ket{d_S}$ in  $\ket{\psi_{RS}}$ are respectively $P(d_Ru_N)=|na^2b|^2$ and $P(d_Rd_S)=|na^2b^2|^2$. Using the classical Bayes' rule, we have
\begin{align}
	P(d_Ru_N|d_Rd_S)= P(d_Rd_S|d_Ru_N) P(d_Ru_N)/P(d_Rd_S)=1.
\end{align} 
Thus, the observation of  $\ket{d_R}\ket{d_S}$ deterministically infers the cause $\ket{d_R}\ket{u_N}$, as exploited to construct the statement (H2).  However, this inference cannot be true because the quantum Bayes' rule implies
\begin{align*}
	\tr_{RS}\left[\mathcal{P}^{U_N^\dag}_{RN|RS}\star \ketbra{d_R}{d_R} \otimes \ketbra{d_S}{d_S}\right] \neq \ketbra{d_R}{d_R} \otimes \ketbra{u_N}{u_N},
\end{align*}
or equivalently, $\mathbb{I} \otimes U_N^\dag \ket{d_R}\ket{d_S}=\ket{d_R}(-b^*\ket{u_N}+a^*\ket{v_N})\neq \ket{d_R}\ket{u_N}$. Here the inverse transformations are $U_{M}^\dag: \mathcal{H}_R \mapsto \mathcal{H}_M$ and $U_{N}^\dag: \mathcal{H}_S \mapsto \mathcal{H}_N$, where
\begin{align*}
	\ket{c_{R/S}} \mapsto a \ket{u_{M/N}} + b \ket{v_{M/N}}, \ \ \ket{d_{R/S}} \mapsto -b^* \ket{u_{M/N}} + a^* \ket{v_{M/N}}.
\end{align*}
It does not satisfy the condition~\eqref{eq:GenEtoC} for deterministic causal inference. One may, at most, claim that the cause $\ket{d_R}\ket{u_N}$ results in the effect $\ket{d_R}\ket{d_S}$ with a probability $|b|^2$. Therefore, the statement made in (H2) is only true probabilistically, and the same applies to the statements (H3)-(H5). Thus, the contradiction as s result of the statements (H0)-(H5) is flawed. 

In fact, following QCR, the observation of the effect $\ket{d_R}\ket{d_S}$ in $RS$ implies that $MN$ should contain the cause $\ket{u_M}\ket{u_N}$ with the probability $|b|^4$, as 
\begin{align*}
U_M^\dag \otimes U_N^\dag	\ket{d_R}\ket{d_S} = (-b^* \ket{u_{M}} + a^* \ket{v_{M}})(-b^* \ket{u_{N}} + a^* \ket{v_{N}}).
\end{align*}
However, it is clear that the cause  $\ket{u_M}\ket{u_N}$ is not present in the state $\ket{\psi_{MN}}$, and it does not lead to a contradiction as such. Because, there are other effects present in $\ket{\psi_{RS}}$ that are in coherent superposition with $\ket{d_R}\ket{d_S}$. Once we consider all these effects in superposition, i.e., the overall effect $\ket{\psi_{RS}}$, and infer the cause by applying $U_M^\dag \otimes U_N^\dag$ or QCR, we see that the overall cause (initial state) does not include $\ket{u_M}\ket{u_N}$, and this is exclusively due to the fact that quantum causes can coherently superpose. Therefore, there is no contradiction once we use quantum Bayes' rule. 

\section{Conclusion \label{sec:Disc}}
We have demonstrated that the inferences (similarly, predictions) following causal relation based on classical Bayes' rule significantly differ from those based on quantum Bayes' rule. The differences in the inferences and predictions using classical and quantum Bayes' rules are due to three main reasons. First,  quantum causes, as well as effects, can coherently superpose. Second, the act of observation of an effect or cause leads to a collapse in the observed system. Third, for bipartite systems, the local effects (also causes) can have strong quantum correlations (like entanglement). Because of that, observing a local effect (cause) on one system may induce a collapse of the other system. But classical Bayes' rule is based on conditional probability, and correspondingly the classical causal relations, completely ignores these aspects that are very particular to quantum mechanics. The use of classical Bayes' rule, in fact, leads to various contradictions in quantum mechanics, and that can again be resolved with the help of quantum Bayes' rule. To demonstrate that, we have considered two cases. One is based on a paradox by Frauchiger and Renner \cite{Frauchiger18}, which claims that quantum mechanics cannot consistently explain the use of itself. We have shown that there is no inconsistency in predictions and measurement inferences if one uses quantum Bayes' rule. The other case, we have considered, is based on the paradox by Hardy \cite{Hardy92, Hardy93}. However, unlike Hardy's paradox, we have assumed that quantum mechanics is non-local and made predictions and inferences based on global measurements. Even in that case,  classical Bayes' rule leads to an irreconcilable contradiction, and it, again, is resolved with the use of quantum Bayes' rule. 

Therefore, we conclude that the classical Bayes' rule is inadequate for quantum mechanics. In order to have consistent predictions and inferences (or causal relation), one must rely on quantum Bayes' rule. In quantum mechanics, classical Bayes' rule is applied in the context of causal inferences and predictions, parameter estimations, state and channel tomography, etc. We anticipate that our findings will have deep implications in these contexts. 

\section*{Acknowledgment}
We thankfully acknowledge fruitful discussions with Renato Renner, Maciej Lewenstein, Antonio Acin, Anderas Winter, Guruprasad Kar, Swapan Rana, Cyril Branciard, Armin Tavakoli, and Markus M\"uller. M.L.B. acknowledges financial supports from ERC AdG NOQIA; Ministerio de Ciencia y Innovation Agencia Estatal de Investigaciones (PGC2018-097027-B-I00/10.13039/501100011033, CEX2019-000910-S/10.13039/501100011033, Plan National FIDEUA PID2019-106901GB-I00, FPI, QUANTERA MAQS PCI2019-111828-2, QUANTERA DYNAMITE PCI2022-132919, Proyectos de I+D+I "Retos Colaboración" QUSPIN RTC2019-007196-7); European Union NextGenerationEU (PRTR); Fundació Cellex; Fundació Mir-Puig; Generalitat de Catalunya (European Social Fund FEDER and CERCA program (AGAUR Grant No. 2017 SGR 134, QuantumCAT \ U16-011424, co-funded by ERDF Operational Program of Catalonia 2014-2020); Barcelona Supercomputing Center MareNostrum (FI-2022-1-0042); EU Horizon 2020 FET-OPEN OPTOlogic (Grant No 899794); National Science Centre, Poland (Symfonia Grant No. 2016/20/W/ST4/00314); MCIN/AEI/10.13039/501100011033 and FSE "El FSE invierte en tu futuro" BES-2017-082828; European Union's Horizon 2020 research and innovation programme under the Marie-Sklodowska-Curie grant agreement No 101029393 (STREDCH) and No 847648 ("La Caixa" Junior Leaders fellowships ID100010434: LCF/BQ/PI19/11690013, LCF/BQ/PI20/11760031, LCF/BQ/PR20/11770012, LCF/BQ/PR21/11840013). M.N.B. acknowledges supports from SERB-DST (CRG/2019/002199), Government of India. 


%

\end{document}